\documentclass[12pt]{article}
\usepackage{latexsym}
\usepackage{epsfig}
\begin{document}

\newcommand{\beq}{\begin{equation}}
\newcommand{\eeq}{\end{equation}}
\newcommand{\bear}{\begin{eqnarray}}
\newcommand{\eear}{\end{eqnarray}}
\newcommand{\half}{{{1}\over{2}}}
\newcommand{\nn}{\nonumber}
\newcommand{\pa}{\partial}

\begin{titlepage}
\begin{flushright} KUL-TF-99/22 \\ gr-qc/9906050
\end{flushright}
\vskip 2.5cm
\begin{center}
{\Large \bf On the (im)possibility of warp bubbles} \\
\vskip 1.5cm
{\bf Chris Van Den Broeck$^\ast$} \\
{\small Physics Division, Starlab Research \\
Boulevard St.--Michel 47, 1040 Brussels, Belgium }
\end{center}
\vskip 4cm
\begin{center}
{\bf Abstract}
\begin{quote}
Various objections against Alcubierre's warp drive geometry are
reviewed. Superluminal warp bubbles seem an unlikely possibility
within the framework of general relativity and quantum field theory,
although subluminal bubbles may still be possible.
\end{quote}
\end{center}
\vfill
\hrule width 5cm
\vskip 2mm
{\small 
$^\ast$ vdbroeck@starlab.net}
\end{titlepage}

\section{Introduction}

Since Alcubierre published his `warp drive' spacetime \cite{Alcubierre}, 
the proposal has been criticized from various viewpoints by a number
of authors \cite{Krasnikov,FordPfenning,Hiscock,Coule}. One of the
problems, concerning the amount of exotic matter needed to support
a warp bubble capable of transporting macroscopic objects 
\cite{FordPfenning}, was partially solved in \cite{Chris}. 
Another objection, claiming a divergence of quantum fluctuations
on a warp drive background, is probably not valid in the general
case \cite{Hiscock}. However, serious problems remain, and as we
will see, it is unlikely that the original ansatz can be modified to 
circumvent all of them, at least for superluminal bubbles.

In the next section, the warp drive geometry is introduced; in the subsequent
section, we review some of the objections that have been raised.
Section 3 deals with the behaviour of quantum
fluctuations on a fixed Alcubierre background. In section 4 we discuss
the unreasonably high energies macroscopic warp bubbles would need.
In section 5 we come to the crucial problem: part of the energy supporting 
the warp drive moves tachyonically. A summary is given in section 6.

\section{The Alcubierre spacetime}

The warp drive metric is
\beq
ds^2 = -dt^2 + (dx-v_s(t)f(r_s)dt)^2 + dy^2 + dz^2, \label{warp}
\eeq
with $r_s=\sqrt{(x-x_s(t))^2+y^2+z^2}$, and $v_s={{dx_s}\over{dt}}$,
where $x_s(t)$ is the path followed by the center of the warp bubble.
The function $f$ has the properties $f(0)=1$ and $f(r_s) \rightarrow 0$ as
$r_s \rightarrow \infty$. $x_s(t)$ is then a timelike geodesic with 
proper time equal to the coordinate time outside the bubble. 
We will assume that $f$ has compact support, which 
we will call the warp bubble. This is a natural assumption since it implies 
that the energy densities associated to the geometry do not stretch all 
the way to spacelike infinity. 

At first sight there is no speed limit, in the sense that if $v_s > 1$, a 
particle 
moving along $x_s(t)$ would be able to outrun a photon moving
in the Minkowskian part of spacetime. This is also characteristic
of traversable wormholes \cite{MorrisThorne}, but unlike wormholes,
the warp drive does not need non--trivial topology. However, it will become 
clear that as soon as the bubble goes superluminal, the geometry (\ref{warp}) 
develops unphysical features, not all of which can be mended by simple 
modifications of the spacetime.

A problem the warp drive has in common with traversable wormholes 
is a violation of the energy conditions of general relativity.
If quantum field theory (QFT) is introduced, this is no longer a crucial
problem; for example, the well--known Casimir effect violates 
the Weak Energy Condition (WEC)\footnote{Helfer et al. \cite{Helfer} argued 
that this would not be true in realizable
experimental set--ups due to the properties of known materials, but their
results do not in principle rule out a Casimir WEC violation.}. As has been 
known for a long time, 
QFT on curved spacetimes indicates that spacetime curvature itself could cause
violations of the energy conditions, and recently a class of 
wormholes was found which would self--stabilize, in the sense that
the negative energy (`exotic matter') densities needed to sustain the wormhole 
geometry would arise from vacuum fluctuations of conformal fields
due to the curvature of the wormhole geometry itself \cite{travworm}.

\section{Quantum fields on an Alcubierre background}

Hiscock \cite{Hiscock} argued that the energy density due to 
fluctuations of conformally coupled quantum fields would diverge
at particle horizons within the bubble, which are present as soon
as $v_s >1 $. The calculation was
only performed for the $1+1$ dimensional version of the warp drive
geometry, but it is reasonable to assume that a similar phenomenon
would occur in four dimensions \cite{Hiscock2}. Hiscock's calculations 
involved a 
coordinate transformation making the warp drive spacetime manifestly 
static. In two dimensions, the geometry turned out to be similar to that
of a 2--dimensional black hole. Calculations of the stress--energy tensor
of a conformal field living on such a background \cite{BHHiscock} indicate
that if the field has reached thermal equilibrium, its temperature at
spacelike infinity must be equal to the Hawking temperature of the 
black hole, otherwise vacuum fluctuations would diverge strongly at 
the horizon. In the case of the constant velocity warp drive, the temperature 
of the horizon would never be equal to that of a field on the 
background, from which a divergence was inferred. However, it is questionable 
that such a divergence would be present if the warp bubble had gone 
superluminal and developed horizons a finite time in the past, a situation 
comparable to that of a newly formed black hole not in thermal equilibrium 
with the field at infinity.

\section{Unreasonably high energies}

Ford and Roman \cite{QI} suggested 
an uncertainty--type principle which places a bound on the extent
to which the WEC is violated by quantum fluctuations of scalar and 
electromagnetic fields: The larger the violation, the shorter the time it 
can last 
for an inertial observer crossing the negative energy region. This so--called 
quantum inequality (QI) can be used as a test for the 
viability of would--be spacetimes allowing superluminal travel. 
By making use of the QI, Ford and Pfenning \cite{FordPfenning} were able to
show that a warp drive with a macroscopically large bubble must contain an 
unphysically large amount of negative energy. This is because the QI 
restricts the bubble wall to be very thin, and for a macroscopic bubble the 
energy is roughly proportional to $R^2/\Delta$, where 
$R$ is a measure for the bubble radius and $\Delta$ for its wall thickness. 
It was shown that a bubble with a radius of 100 meters would require a 
total negative energy of at least
\beq
E \simeq - 6.2 \times 10^{62} v_s \,\, \mbox{kg}, 
\eeq
which, for $v_s \simeq 1$, is ten orders of magnitude bigger than the total 
positive mass of the entire visible Universe.

In \cite{Chris}, it was shown that this number is very much dependent 
on the details of the geometry. The total energy can be reduced
dramatically by keeping the surface area 
of the warp bubble itself microscopically 
small, while at the same time expanding the spatial volume inside the 
bubble. The most natural way to do this is the following:
\beq
ds^2 = - dt^2 + B^2(r_s) [(dx - v_s(t) f(r_s) dt)^2 + dy^2 + dz^2].
\label{metric}
\eeq
$B(r_s)$ is a 
twice differentiable function such that, for some $\tilde{R}$ and 
$\tilde{\Delta}$,
\bear
B(r_s)=1+\alpha & \mbox{for} & r_s <\tilde{R}, \nn\\
1<B(r_s) \leq 1+\alpha & \mbox{for} & \tilde{R} \leq r_s < \tilde{R} 
+ \tilde{\Delta}, \nn\\
B(r_s)=1 & \mbox{for} & \tilde{R} + \tilde{\Delta} \leq r_s,
\eear
where $\alpha$ will in general be a very large constant; $1+\alpha$ is the
factor by which space is expanded. For $f$ one chooses a function
with the properties 
\bear
f(r_s)=1 & \mbox{for} & r_s < R, \nn\\
0 < f(r_s) \leq 1 & \mbox{for} & R \leq r_s < R + \Delta, \nn\\
f(r_s)=0 & \mbox{for} & R + \Delta \leq r_s, \nn
\eear
where $R > \tilde{R} + \tilde{\Delta}$.

A spatial slice of the geometry one gets in this way can be easily visualized
in the `rubber membrane' picture. A small Alcubierre bubble surrounds a neck 
leading to a `pocket' with a large internal volume, with a flat region in the 
middle. It is easily calculated that the center $r_s=0$ of the pocket
will move on a timelike geodesic with proper time $t$.

Using this scheme, the required total energy can be reduced to stellar 
magnitude, in such a way that the QI is satisfied. On the other hand,
the energy densities are still unreasonably large, and the spacetime
has structure with sizes only a few orders of magnitude 
above the Planck scale.

\section{Energy moving locally faster than light}

The most problematic feature of the warp drive geometry is 
the behaviour of the negative energy densities
in the warp bubble wall \cite{Krasnikov,Coule}. If the Alcubierre spacetime 
is taken literally, part of the exotic matter will have to move 
superluminally {\sl with respect to the local lightcone}. 
It is easy to see that all exotic matter outside some surface surrounding
the center (let us call this the critical surface), will move in a
spacelike direction. For $v_s>1$, there has to be exotic matter
outside the critical surface, since the function $f$ must reach the value $0$ 
for some 
$r_s$ (which, of course, can be infinity), and the negative energy density 
is proportional to $\left({{df}\over{dr_s}}\right)^2$ for an 
`Eulerian observer'
\cite{Alcubierre}. As noted in \cite{Coule}, the Alcubierre spacetime
is an example of what can happen when the Einstein equations are run
in the `wrong' direction, first specifying a metric, then calculating
the associated energy--momentum tensor.

The problem of tachyonic motion can be interpreted as meaning that
part of the necessary exotic matter is not able to keep up with 
the rest of the bubble: if one would try to make a warp bubble go 
superluminal, the outer shell would be left behind, destroying the
warp effect.

It is conceivable that the problem can be circumvented, for example by
letting the distribution of exotic matter expand into a `tail' in the back. 
It may
be possible to do this in a way compatible with both the QI and the Quantum 
Interest Conjecture introduced in \cite{interest}, which states that
a pulse of negative energy must always be followed by a larger pulse
of positive energy. However, it is unlikely that one can get rid
of tachyonic motion of the exotic matter without 
introducing a naked curvature singularity in the front of the bubble 
\cite{unpublished}.

\section{Summary}

It would seem that the main problems of the warp drive can not be
solved without retaining some unphysical features or introducing
new ones, such as high energy densities, curvature radii of the order of the 
Planck length, and naked singularities. We have limited the
discussion to superluminal warp bubbles. Subluminal bubbles 
are still an open possibility, and it is not inconceivable that microscopic 
ones might even occur naturally. Due to the absence of horizons, potential
problems due to diverging vacuum fluctuations will not arise, and there will
be no tachyonic motion of exotic matter. Possibly the geometry
can be chosen in such a way that the necessary negative energy densities 
are partly
supplied by the changes in vacuum fluctuations induced by the curvature, as 
in \cite{travworm}. An interesting question is whether one can 
construct a spacetime similar to the subluminal warp drive which avoids 
negative energy densities altogether, but for this no ansatz is available
at the present time.

\medskip
\section*{Acknowledgements}

I have benefited from the interesting comments of M. Alcubierre, 
D.H. Coule, and S.V. Krasnikov. It is a pleasure to thank the European Office 
for Aerospace Research and Development for financial support.

\end{document}